\documentstyle[12pt]{article}
\newcommand{\Prob}[1]{\mathbin{\bf{P}}\left\{#1\right\}}

\newcommand{\textover}[2]{{\textstyle{#1\over#2}}} %force a `fraction'
\newcommand{\greekvec}[1]{{{#1}}} %this should embolden Greek characters!
\newcommand{\greekmatr}[1]{{{#1}}} %this should embolden Greek characters!
\renewcommand{\vec}[1]{{\bf{#1}}}
\newcommand{\matr}[1]{{\bf{#1}}}

\def\dotsc{\ldots}
\def\ie{{\em i.e.}}

\date{}

\begin{document}

\title{\LARGE\sf Large Fluctuations in Stochastically Perturbed Nonlinear
Systems: Applications in Computing}
\author{
Robert S. Maier\thanks{
Supported in part by the National Science Foundation
under grant NCR-90-16211.}\\
{\small \em Department of Mathematics}\\
{\small \em University of Arizona}\\
{\small \em Tucson, Arizona 85721, U.S.A.}
}

\maketitle

\section{Introduction}
\label{sec:intro}
Nonlinear dynamical systems often display complex behavior.  In~this
lecture I~shall review the behavior of {\em stochastically perturbed\/}
dynamical systems, which is a field of its~own.  I~shall use this as an
opportunity to discuss applications to computer science, though
applications to statistical physics, chemical physics, and elsewhere in the
sciences are also numerous.

If a deterministic dynamical system has an attractor, by~definition the
system state approaches the attractor in the long-time limit.  But if the
system is regularly subjected to small stochastic fluctuations (random
kicks, or~noise) this approach will only be approximate.  In~the long-time
limit the system state will typically be specified by a probability
distribution (a~`noisy attractor') centered on the attractor proper.
In~the limit as the noise strength tends to zero, this distribution will
converge to the attractor.

Even if the system has a single globally stable point as its only
attractor, one can pose an interesting question: What is the probability,
if~the noise strength is very small, of~finding the system in a specified
state macroscopically distant from the attractor?  How long must one wait
before this occurs?  If~the system has more than a single stable state,
each with its own basin of attraction, one can similarly ask for the
timescale on~which transitions between the two basins occur.  Such
questions are really questions about the character of the extreme tail of
the noisy attractor, and can be answered only by quantifying the
probability of {\em large fluctuations\/} of the system.  The mathematical
field dealing with such matters is known as large deviation
theory~\cite{VF,Varadhan}.

In scientific applications one would usually like to know not only how
frequently atypical fluctuations occur, but also along which trajectory the
system state moves during transitions from one stable state to another.
It~turns out that in~most stochastically perturbed dynamical systems
a~single trajectory in the system state space, or at~most a discrete set,
is singled out in the limit of weak noise as by~far the most likely.

This phenomenon has long been known to chemical and statistical physicists,
but its importance in other fields which make use of stochastic modelling,
such as ecology and evolutionary biology, has only recently become
clear~\cite{Lande85,Newman85}.  In~chemical physics the most likely
transition trajectory is interpreted as a~reaction pathway, since chemical
reactions are modelled as transitions from a~metastable state to a~more
stable state~\cite{vanKampen81}.  But the mathematical approach I~shall
sketch is much more general: the dynamical system can be continuous or
discrete, and the system dynamics need not obey detailed balance.  Some of
the strongest results on systems without detailed balance have only
recently been obtained~\cite{MaierA,MaierB}.  The system can even be {\em
distributed\/}, with nontrivial spatial extent; this includes stochastic
cellular automata, and systems specified by stochastic partial differential
equations rather than stochastic ordinary differential equations.

The quasi-deterministic phenomena (optimal trajectories, well-defined
reaction pathways, {\em etc.}) which arise in stochastically perturbed
dynamical systems can be viewed as {\em emergent\/}.  They are determined
by the stochastic dynamics, but in a~rather complicated way, and they
manifest themselves only in the weak-noise limit.  Their appearance in
computer science applications is not well known; I~hope the two examples
treated in this lecture will correct that.  Attempts have recently been
made to interpret the behavior of computers, or interacting networks of
computers, in dynamical system terms or even ecological
terms~\cite{Huberman88}.  But stochasticity is, I~think, a~crucial part of
any such interpretation.

\section{A Simple Stochastic Model: {ALOHA}net}
As a first example drawn from computer science, consider a stochastic model
which attempts to capture the essential features of a large number of
computers communicating with each other across a data network, such as an
Ethernet.  The model will be idealized, but it will be typical~of (``in~the
same universality class~as'') models in which a large number of agents
share occasional access to a single resource.  Here the resource will be
the network bus: the~ether, which only one computer can use at~a~time.

You are no doubt familiar with such application programs as {\tt telnet}
and {\tt ftp}, which allow a user of one machine to communicate with
another.  Behind the scenes (``at~a lower protocol layer,''
in~telecommunications jargon) these programs work as
follows~\cite{Stallings88}.  A~connection between two computers consists of
a stream of data packets, each typically containing between $10$ and~$10^3$
bytes.  (A~data packet is simply a train of square waves.)  An~interactive
login program like {\tt telnet} normally transmits a~packet whenever the
user presses a~key; the packet contains the typed character.  Less
interactive programs like {\tt ftp}, which transfers whole files, employ
larger packets.  There is a scheme known as TCP/IP (Transmission Control
Protocol/Internet Protocol) for specifying the destination of packets, and
for keeping the two communicating computers synchronized.  This last task
may involve the transmission of additional packets.

Let us suppose that a computer is making substantial use of the network:
several users are running {\tt ftp} simultaneously, for example.  In~such
a~situation a~statistical treatment is possible.  In the context of a
particular stochastic model, it~is possible to estimate mean network usage,
and the probability that data packets are transmitted successfully.  That
is what I~shall now~do.

A slight digression is necessary on the issue of {\em successful\/}
transmission.  Ethernet, besides being a tradename, is a multiaccess
protocol: a~scheme for sharing access to the cable connecting two or more
computers.  Normally when a~computer wishes to transmit a~packet, it~does
so immediately.  It~is possible therefore for two machines to transmit
colliding packets, in~which case both packets are corrupted:
the~information in both is lost.  The Ethernet protocol (a~CSMA/CD [Carrier
Sense Multiple Access/Collision Detect] protocol) embodies a~heuristic for
minimizing the probability of collisions, \ie, of~unsucessful transmissions.

A description of the protocol may be found in the book of Bertsekas and
Gallager~\cite{Bertsekas87}.  On~grounds of simplicity I~shall model
a~conceptually similar but simpler protocol known as ALOHAnet.  ALOHAnet
was one of several Ethernet precursors, developed at the University of
Hawaii during the 1970's.  Although it has long since been superseded,
it~lives~on in~the form of a tractable mathematical model.  The stochastic
ALOHAnet model is a discrete-time model or Markov chain, unlike the
continuous-time models which must be employed in the performance analysis
of real-world Ethernets.  The following description is
standard~\cite{Gunther2,Lim87,Maier7b,Nelson}.

Suppose that $N$~computers are attached to the network; $N$~will eventually
be taken to infinity, yielding a continuum limit which (if~proper scaling
is imposed) can be viewed as a weak-noise limit.  At each integer time
$j=1,2,3,\dotsc$ a~packet of data originates with probability~$p_0$ on each
computer not currently blocked.  When is a~computer blocked?  When
a~previously generated packet has failed to be transmitted successfully,
and the packet is awaiting retransmission.

Newly generated packets are always transmitted immediately, but of course
they may collide with packets transmitted by other computers at the same
integer time.  Such collisions are immediately detected, and each of the
transmitting computers enters a blocked state (if~it was not blocked
already).  While in the blocked state, at~each subsequent integer time a
computer will attempt a retransmission with probability~$p_1$.  In~other
words each of the blocked computers backs off a~random amount of time, and
tries again to transmit its packet.  The backoff time is geometrically
distributed, with parameter~$p_1$.  This random backoff policy facilitates
the breaking of the deadlock: if~the blocked computers each backed off a
{\em fixed\/} amount of time, they would simply run into each other again.

This ALOHAnet model has only three parameters: $p_0$, $p_1$, and~$N$.
If~$y_j$ is the number of computers blocked at time~$j$, then
$y_1,y_2,y_3\dotsc$ is a Markov chain on the discrete state space
$\{0,1,2,\dotsc,N\}$.  Let us analyse this Markov chain.

At any time~$j$, the number of retransmitted packets is binomially
distributed, with parameters~$p_1$ and~$y_j$.  Similarly, the number of
newly generated (and transmitted) packets is binomially distributed with
parameters $p_0$ and~$N-y_j$.  If~$X_1$ and~$X_0$ denote these two random
variables, the total number of packets transmitted at integer time~$j$ is
$X_1+X_0$, and
\begin{equation}
\label{eq:cases}
\xi \equiv y_{j+1}-y_j =
\cases{-1,&if $X_0=0$, $X_1=1$;\cr
       X_0,&if $X_0+X_1>1$;\cr
       0,&otherwise.\cr}
\end{equation}
$y_j$~will decrease by~$1$ if a previously unsuccessfully transmitted
packet (and only that packet) is retransmitted.  It~will increase by~$X_0$
in~the event of a collision, and so~forth.  From~(\ref{eq:cases}), it~is
easy to work out the density of the random variable~$\greekvec\xi\equiv
\Delta y$.

Since we wish to construct a continuum large-$N$ limit we define the
normalized network state~$x$ at any time to be~$y/N$, the fraction of
computers that are currently blocked.  Necessarily $0\le x\le 1$.  Besides
scaling the state space in~this way, we~scale time by defining normalized
time~$t$ to equal~$j/N$, so~that~$x$, if~viewed as a function of~$t$, jumps
at $t=1/N,2/N,\dotsc$ by a random quantity~$N^{-1}\xi$.  The density of the
random variable~$\xi$ is specified by the current normalized state~$x$;
we~write $\xi$ as~$\xi(x)$ to make this clear.

To get a nontrivial large-$N$ limit we need to scale the probabilities
$p_0$ and~$p_1$ as well; we~take $p_0=q_0/N$ and~$p_1=q_1/N$, for some
$N$-independent $q_0$~and~$q_1$.  So~$q_0x$ is the expected number of newly
generated packets, and $q_1(1-x)$ the expected number of retransmitted
packets, at~any specified normalized time~$j/N$.  It~is an easy exercise to
verify that in the large-$N$ limit
\begin{equation}
\label{eq:xi}
\langle\xi(x)\rangle = q_0(1-x) -[q_0(1-x)+q_1x]
\exp\left[-q_0(1-x)-q_1x\right]
\end{equation}
is the expected change in the number of blocked computers, at~any specified
time~$j/N$.  The formula~(\ref{eq:xi}) gives us an explicit expression for
$\langle\Delta x\rangle$, the mean amount by which the normalized state~$x$
changes at any specified time~$j/N$; it~is simply
$N^{-1}\langle\xi(x)\rangle$.  So~in the large-$N$ limit the dynamics of
our network model are {\em in~expectation\/} completely specified
by~(\ref{eq:xi}).

We can now see how the ALOHAnet model can be viewed as a stochastically
perturbed dynamical system.  In~expectation, the large-$N$ ALOHAnet model
looks very like a one-dimensional dynamical system
\begin{equation}
\dot x(t) = \langle\xi(x)\rangle.
\end{equation}
defined on the closed interval $[0,1]$.  Such an associated deterministic
dynamical system is called a~{\em fluid approximation\/} by network
performance analysts.  Although (as~we shall see) it~cannot answer the
questions about large fluctuations in~which we~are interested, the fluid
approximation says quite a bit about the stability of the network.
In~Fig.~1, the drift field~$\langle\xi(x)\rangle$ is plotted as a function
of~$x$, for $q_0=0.43$ and~$q_1=5.0$ (parameter values originally chosen by
G\"unther and Shaw~\cite{Gunther2}).  It~is clear that for this choice of
parameters the system has two point attractors: $x_0\approx0.150$
and~$x_1\approx0.879$.  Each has its own basin of attraction, and in~the
fluid approximation the network state flows deterministically to one or the
other.  The two attractors are interpreted as~follows.  Networks,
in~particular heavily loaded networks, are prone to {\em congestion\/}, and
the two attractors are respectively a low-congestion and a high-congestion
state.

The presence of more than a single attractor, for certain parameter values,
is an unfortunate feature of the ALOHAnet protocol.  If~at time zero all
computers begin unblocked, with these parameter values the fraction of
blocked computers will swiftly rise to~$\approx0.150$.  If~on the other
hand at time zero the computers all begin in the blocked state, the
fraction will decrease to~$\approx0.879$ and no further.  In~the latter
case very few packets are successfully transmitted or retransmitted, since
the probability of more than a single computer transmitting a packet is
always very high.  (Since $q_1=5.0$, when $x\approx1$ about 5~computers,
on~average, attempt to retransmit a packet at each time~$j/N$.)  The
ALOHAnet protocol makes no provision for breaking the deadlock by sharing
the network in a sequential or round-robin fashion: in~the event of extreme
congestion, the computers get in each others'~way.

The appearance of more than a single point attractor is actually a bit
atypical; it~will occur only for certain values of the scaled parameters.
(See~Fig.~2.)  The $(q_0,q_1)$-plane is divided into two regions:
a~monostable (one-attractor) region, and a bistable (two-attractor) region.
The equilibrium blocking fraction is a single-valued function
of~$(q_0,q_1)$ in the former region, and a double-valued function in the
latter.  Nelson~\cite{Nelson84} has shown that this phenomenon, which is so
suggestive of statistical-mechanical critical behavior, generalizes
naturally to multidimensional parameter spaces.  The Ethernet protocol
modifies the packet retransmission probability each time an unsuccessful
retransmission occurs, so a more realistic ALOHAnet model would be
specified by a vector $(p_0,p_1,p_2,\dots)$ of~probabilities,
with $p_k$,\ $k\ge1$, the probability of transmitting a packet which has failed
to be
successfully transmitted exactly $k$~times.  The corresponding normalized
system state would be a vector $(x^{(1)},x^{(2)},\dotsc)$ of blocking
fractions: $x^{(k)}$,\ $k\ge1$, would be the fraction of computers which
are blocked and which have failed to transmit a stored packet exactly
$k$~times.  The analogue of Fig.~2 would be a multidimensional phase
diagram, some regions in which would be characterized by the presence of
multiple point attractors in the multidimensional normalized state space.

The preceding treatment has been entirely in the context of the
deterministic fluid approximation.  The network state does not actually
evolve deterministically, except in expectation.  The expected increment
$\langle\Delta x\rangle$ equals $N^{-1}\langle\xi(x)\rangle$, but the
standard deviation of~$\Delta x$ is also proportional to~$N^{-1}$.  $\Delta
x$~equals $\langle\Delta x\rangle$ plus $\Delta x -\langle\Delta x\rangle$,
and the latter term can be viewed as a stochastic perturbation superimposed
on the dynamical system.  These stochastic perturbations will broaden the
point attractors into noisy attractors, and occasionally induce transitions
between them.

These transitions are of considerable practical interest, since they are
sudden changes in network congestion.  A~heavily loaded network can
suddenly shift from a low-congestion state to a high-congestion state,
in~which almost no packets are transmitted successfully.  (This has rather
drastic effects on the computers attached to the network!)  But to model
such transitions, a~fully stochastic treatment is necessary.

\section{Wentzell-Freidlin Theory}
The techniques employed to estimate the transition time between metastable
states, and in~general to estimate the probability of unlikely events in
the weak-noise limit, go~under the name of Wentzell-Freidlin
theory~\cite{Varadhan}.  Wentzell-Freidlin theory is simply the large
deviation theory of stochastically perturbed dynamical systems.  Many
results in this area are due to physicists and
chemists~\cite{Hanggi90,Schuss80,vanKampen81}, but Wentzell and Freidlin
were the first to put the subject on a sound mathematical
footing~\cite{VF,Wentzell76}.  I~shall summarize their main results, and
extensions.

Consider a multidimensional random process ${\vec x}(t)$ similar to the
normalized ALOHAnet process.  ${\vec x}(t)$~is assumed to jump at times
$t=N^{-1},2N^{-1},3N^{-1},\dotsc$, and the jump magnitude is $N^{-1}$ times
a random vector whose distribution depends on the current state~${\vec x}$.
We~write this random vector as~${\greekvec\xi}({\vec x})$, so $\Delta {\vec
x}=N^{-1}{\greekvec\xi}({\vec x})$.  The $N\to\infty$ limit will be a
weak-noise limit.

This random process strongly resembles a diffusion process with drift.
In~fact the expected drift velocity at any point~${\vec x}$ is ${\vec
u}({\vec x})\equiv\langle{\greekvec\xi}({\vec x})\rangle$, and the
diffusion tensor is $N^{-1}$~times $D_{ij}({\vec x})\equiv{\mathop{\rm
Cov}}(\xi_i({\vec x}),\xi_j({\vec x}))$, the covariance matrix of the
components of~${\greekvec\xi}({\vec x})$.  A~continuous-time diffusion
process~${\vec x}(t)$ with these parameters would satisfy the stochastic
differential equation
\begin{equation}
\label{eq:stochDE}
d x_i(t) = u_i({\vec x}(t)) +
\sum_j\frac{\sigma_{ij}({\vec x}(t))}{\sqrt{N}} dw_j(t)
\end{equation}
where $d{\vec w}(t)$ is white noise, and the
tensor~${\greekmatr\sigma}=(\sigma_{ij})$ is related to the tensor~${\matr
D}=(D_{ij})$ by ${\matr D}={\greekmatr\sigma}{\greekmatr\sigma}^t$.  But this
continuous-time `diffusive approximation' to the underlying jump process is
not especially useful for our purposes: the large fluctuations of the jump
process turn out to depend crucially on the higher moments
of~${\greekvec\xi}({\vec x})$.

Suppose that ${\vec x}_0$ is an attractor for the expected drift
field~${\vec u}({\vec x})$.  Then in expectation ${\vec x}(t)$~will tend to
flow toward~${\vec x}_0$ if~it begins in the basin of attraction of~${\vec
x}_0$.  Thereafter, ${\vec x}(t)$~will tend to wander near~${\vec x}_0$ for
a long time.  But statistical fluctuations of all magnitudes will occur;
the stochastic perturbations $N^{-1}[{\greekvec\xi}({\vec x})-{\vec
u}({\vec x})]$ will eventually push~${\vec x}$ outside any specified
region~$U$ surrounding~${\vec x}_0$.  In~other words, the noise will
eventually overcome the drift.

Since the effective diffusion coefficient decays as~$N^{-1}$, one expects
that the time to exit any specified region~$U$ grows (in~expectation)
exponentially in~$N$.  That is correct, and Wentzell-Freidlin theory
provides a technique for computing the asymptotic exponential growth rate.
This will of course depend on the choice of~$U$.  In~most applications
$U$~is the entire basin of attraction of the attractor~${\vec x}_0$, though
a smaller region could be chosen.

The technique is as follows.  According to theory the expected exit
time~$\langle t_{\rm exit}\rangle$ has weak-noise asymptotics
\begin{equation}
\langle t_{\rm exit}\rangle \sim \exp(N{\cal S}_0), \qquad N\to\infty
\end{equation}
where
\begin{equation}
{\cal S}_0 = \inf \int L({\vec x}(t),\dot{\vec x}(t))\,dt
\end{equation}
is a {\em minimum action\/} for exiting trajectories.  The infimum is taken
over all trajectories~${\vec x}(t)$ which begin at~${\vec x}_0$ and
terminate on the boundary of~$U$.  The transit time is left unspecified.
Here $L({\vec x},\dot{\vec x})$ is a Lagrangian function, dual to a
Hamiltonian or energy function constructed from the distribution
of~${\greekvec\xi}({\vec x})$ by the formula
\begin{equation}
\label{eq:H}
H({\vec x},{\vec p}) = \log \langle \exp({\vec p}\cdot{\greekvec\xi}({\vec
x}))\rangle.
\end{equation}
It is clear that the higher moments of~${\greekvec\xi}({\vec x})$ enter
into the computation of the function~$H$.  $H({\vec x},\cdot)$~is in fact
the cumulant generating function of the random
variable~${\greekvec\xi}({\vec x})$.

The sudden appearance of a classical Hamiltonian and its dual Lagrangian is
quite remarkable.  They are not mere mathematical auxiliaries.  The
trajectory~${\vec x}^*(t)$ minimizing the action (it~usually exists, and is
unique) is interpreted as the {\em most probable exit path\/} (MPEP) in the
limit of weak noise.  It~is not difficult to check, using standard methods
of classical mechanics, that the optimization of the action over transit
times yields an MPEP which is a {\em classical trajectory of zero
energy\/}.  So~the `momentum'~${\vec p}$, which has no direct physical
interpretation, as a function of position~${\vec x}$ along the MPEP must
satisfy
\begin{equation}
\langle \exp({\vec p}\cdot{\greekvec\xi}({\vec x}))\rangle = 1.
\end{equation}
If the state space is one-dimensional, this zero-energy constraint alone
will determine the MPEP.

The MPEP ${\vec x}^*$ is not only a most probable exit path: it~is also an
exit path of least resistance.  Although ${\vec x}(t)$ will remain in~$U$
for an exponentially long time, it~will fluctuate out along the MPEP (and
in other directions) an~exponentially large number of times before the MPEP
is traversed in~full and $U$~is exited.  The final fluctuation will
follow~${\vec x}^*$ quite closely in the large-$N$ limit.  One can view the
equilibrium distribution of the system state~${\vec x}$ (the~noisy
attractor) as being concentrated near~${\vec x}_0$, but having a tube-like
protuberance stretching out toward the boundary of~$U$ along the
trajectory~${\vec x}^*$.  In~the large-$N$ limit the tube is exponentially
suppressed, and the noisy attractor converges to the point attractor~${\vec
x}_0$.

$\langle t_{\rm exit} \rangle$ grows exponentially in~$N$, but the limiting
{\em distribution\/} of~$t_{\rm exit}$ has not yet been specified.
It~turns out to be an exponential distribution.  This is very typical of
weak-noise escape problems, where the probability of any single escape
attempt is small.  (The same exponential distribution is seen in
radioactive decay.)

${\cal S}_0$, the weak-noise growth rate of the expected exit time, can be
viewed as a {\em barrier height\/}: a~measure of how hard it~is to overcome
the drift driving~${\vec x}$ toward~${\vec x}_0$ and away from the boundary
of~$U$.  In~fact the Wentzell-Freidlin framework, if~extended to
conservative continuous-time processes described by~(\ref{eq:stochDE}),
yields the familiar Arrhenius law for the growth of the exit time in the
limit of weak noise.  For such systems ${\cal S}_0$~is simply the height of
the potential barrier surrounding the attractor.

What is not clear from the Wentzell-Freidlin treatment (and~is still not
{\em rigorously\/} clear, though numerous nonrigorous results have been
obtained~\cite{MaierA,Matkowsky84,Naeh90}) is~the subdominant large-$N$
asymptotics of~$\langle t_{\rm exit} \rangle$.  In~general one expects
\begin{equation}
\label{eq:fullas}
\langle t_{\rm exit} \rangle \sim CN^{\alpha} \exp(N{\cal S}_0), \qquad
N\to\infty,
\end{equation}
for some constants $C$~and~$\alpha$, but Wentzell-Freidlin theory yields
only the exponential growth rate~${\cal S}_0$.  The pre-exponential factor
in~(\ref{eq:fullas}) remains to be determined.

The current status of the prefactor problem can be summed~up as~follows.
If~$U$ is taken to be the entire basin of attraction of~${\vec x}_0$,
$\alpha$~is typically zero, and $C$~can be obtained by a method of matched
asymptotic expansions, \ie, a~method of systematically approximating the
equilibrium distribution of~${\vec x}$.  However in multidimensional models
there is an entire zoo of possible pathologies, including the appearance of
caustics and other singular curves in the state
space~\cite{Chinarov93,MaierA,MaierB}, which can induce a nonzero~$\alpha$
and/or hinder a straightforward computation of~$C$.  This is the case,
at~least, for continuous-time diffusion processes defined by stochastic
differential equations.  The situation for jump processes is expected to be
similar.

\section{Applying the Theory}
Wentzell-Freidlin theory, with extensions, can be applied to the stochastic
ALOHAnet model, and to other stochastically perturbed dynamical systems
arising in computer science.  The quantity most readily computed is~${\cal
S}_0$, the exponential growth rate in the weak-noise limit of the expected
time before the system leaves a specified region surrounding a point
attractor in the system state space.  Recall that in the ALOHAnet model
this region is the basin of attraction; departure from~it signals a drastic
change in network congestion.

If the system state space is one-dimensional, as in the ALOHAnet model, the
classical-mechanical interpretation of~${\cal S}_0$ facilitates its
computation.  ${\cal S}_0$~is always the action of a zero-energy
trajectory, with energy as a function of position and momentum given by the
formula~(\ref{eq:H}).  This Hamiltonian is a convex function of~${\vec p}$
at fixed~${\vec x}$, so~if the state space is one-dimensional
(and~$\langle{\greekvec\xi}({\vec x})\rangle\neq{\vec0}$, which~will always
be the case within the basin of attraction) the equation $H({\vec x},{\vec
p})=0$ will have only two solutions for ${\vec p}={\vec p}({\vec x})$.  One
of these is ${\vec p}\equiv{\vec 0}$, which is unphysical.  This solution
is unphysical because if~${{\vec p}={\vec 0}}$
\begin{equation}
\dot{\vec x} = \frac{\partial H}{\partial{\vec p}}
=\frac{ \langle{\greekvec\xi}({\vec x}) \exp({\vec
       p}\cdot{\greekvec\xi}({\vec x}))\rangle}
{ \langle \exp({\vec
       p}\cdot{\greekvec\xi}({\vec x}))\rangle}
=\langle {\greekvec\xi}({\vec x}) \rangle
\end{equation}
and the ${\vec p}\equiv{\vec 0}$ trajectory simply follows the mean drift,
which points {\em toward\/} the attractor rather than away.  The MPEP must
be a classical trajectory emanating from the attractor, so in a
one-dimensional system it is uniquely characterized by the condition that
${\vec p}={\vec p}({\vec x})$ be the nonzero solution of $H({\vec x},{\vec
p})=0$.  Actually there are two such trajectories, one emanating to either
side of the attractor; the true MPEP will be the one with lesser action.

In general to compute~${\cal S}_0$, even in higher-dimensional models one
needs only the MPEP and the momentum as a function of position along~it.
This is because the action of any zero-energy classical trajectory may be
written as a line integral of the momentum, so~that
\begin{equation}
\label{eq:easy}
{\cal S}_0 = \int {\vec p}({\vec x})\cdot d{\vec x}
\end{equation}
the integral being taken along the MPEP from the attractor to the boundary
of the region.  But only in one-dimensional models is~(\ref{eq:easy}) easy
to apply.  In~$d$-dimensional models merely finding the MPEP requires an
optimization over the $(d-1)$-dimensional family of zero-energy
trajectories extending to the boundary.  Except in models with symmetry,
this optimization must usually be performed numerically.

\subsection{The ALOHAnet Application}
In the ALOHAnet model, the expected drift $\langle\xi(x)\rangle$ as a
function of normalized network state~$x$ is given by~(\ref{eq:xi}).  But
to~study large fluctuations, and compute the MPEP, one needs the
Wentzell-Freidlin Hamiltonian $\log\langle\exp({ p}{\xi}({ x}))\rangle$.
In~the large-$N$ limit the random variables $X_1$~and~$X_0$, in~terms of
which $\xi$~is expressed by~(\ref{eq:cases}), become respectively a Poisson
random variable with parameter~$q_1x$ and a Poisson random variable with
parameter~$q_0(1-x)$.  A~bit of computation yields
\begin{equation}
H(x,p) =
\log \left[e^{q_0(1-x)(e^p-1)}
+q_0(1-x)e^{-q_0-q_1x}(1-e^p)
+q_1xe^{-q_0-q_1x}(e^{-p}-1)\right]
\end{equation}
as the Hamiltonian.

If the parameters $q_0$ and $q_1$ are known, it~is easy to compute the
momentum $p=p(x)$ along the MPEP, by numerically solving for the nonzero
solution of the implicit equation $H(x,p(x))=0$.  But the MPEP, and
hence~${\cal S}_0$, will depend on the choice of basin of attractor.  With
the parameter values $q_0=0.43$ and~$q_1=5.0$ of Fig.~1, the two attractors
$x_0\approx0.150$ and~$x_1\approx0.879$ have respective basins of
attraction $[0,x_c)$ and~$(x_c,1]$, with $x_c\approx0.278$ the intermediate
repellor.  MPEPs extend from $x_0$~to~$x_c$, and from $x_1$~to~$x_c$.
Numerical integration of~$p(x)$ gives
\begin{eqnarray}
{\cal S}_0[x_0\to x_c] &\approx &0.00177 \\
{\cal S}_0[x_1\to x_c] &\approx &0.014
\end{eqnarray}
as the growth rates of the expected transition times.

We~see that for the stochastically modelled ALOHAnet, in~the large-$N$
limit a reduced description is appropriate.  Asymptotically, it~becomes a
{\em two-state process\/}.  The network is either in a low-congestion state
(the~basin of attraction of~$x_0$) or a high-congestion state (the~basin of
attraction of~$x_1$), and the transition rates between them
(the~reciprocals of the expected transition times) display exponential
falloffs
\begin{equation}
\exp \left(-N{\cal S}_0[x_0\to x_c]\right), \qquad
\exp \left(-N{\cal S}_0[x_1\to x_c]\right)
\end{equation}
respectively.  With the above choice of parameters, for
reasonable-sized~$N$ the latter transition rate is much smaller than the
former.  The network, once congestion has interfered with the proper
performance of the backoff algorithm, gets `stuck' for potentially a
long time.  This is clearly not a good choice of network parameters!

In a real-world $N$-computer ALOHAnet implementation, $q_0$~would be the
total network load, and would be determined by the level of interprocessor
computing taking place on the network.  The backoff parameter~$q_1=Np_1$
however would probably be fixed, with $p_1$ hardcoded in a data
communications chip installed in each computer.  So~the Wentzell-Freidlin
approach could be employed to determine the likelihood, as~a function of
network load, of~irreversible (or~all but irreversible) congestion
occurring.

Of course the bistability of the system is itself a function of
$q_0$~and~$q_1$.  As~noted, for many values of the parameters the network
is monostable: there is only a single attractor, which may be characterized
by a comparatively low level of congestion.  For such a network one could
compute an action~${\cal S}_0$ for any specified maximum tolerable
congestion level.  The associated optimal (\ie,~most probable) approach
path would be computed much as the MPEP is computed in the bistable case.

\subsection{A Colliding Stacks Application}
There have been several applications of large deviation theory to the
stochastic modelling of {\em dynamic data
structures\/}~\cite{Louchardprime,Maier6,Maier3}.  The memory usage of a
program or programs being executed by a computer can be modelled as a
discrete-time jump process.  In~many cases this process may be viewed as a
finite-dimensional dynamical system, subject to small stochastic
perturbations.  Of~interest is the amount of time expected to elapse before
a particularly large fluctuation away from a deterministic point attractor
occurs.  This would correspond, in~real-world terms, to~an atypical string
of memory allocations leading to an exhaustion of memory.

The following two-dimensional `colliding stacks' model was first studied by
Flajolet~\cite{Flajolet}, having been first suggested by Knuth.  Suppose
that $N$~cells of memory, arranged in a linear array, are available for use
by two programs.  Suppose that at~any given time, the programs will require
$y^{(1)}$~and~$y^{(2)}$ cells of memory respectively.  It~will be most
efficient
for them to employ respectively the first~$y^{(1)}$ and the last~$y^{(2)}$
cells of
the array, so~as to avoid contention for memory.  It~is necessary that
$y^{(1)}+y^{(2)}\le N$; if~this inequality becomes an equality, the two-program
system runs out of memory.

A natural model for the evolution of $y^{(1)}$~and~$y^{(2)}$ is as~follows.
At~any integer time $j=1,2,3,\dotsc$, there are four possibilities:
$y^{(1)}$~may increase by~$1$, $y^{(1)}$~may decrease by~$1$, $y^{(2)}$~may
increase by~$1$, and $y^{(2)}$~may decrease by~$1$.  These are assigned
probabilities $p/2$, $(1-p)/2$, $p/2$, $(1-p)/2$, for $p$ the probability
of a net increase in memory usage.  Let us take $0<p<\frac12$, so~that
deallocations of memory are more likely than new allocations.  (Note that
if $y^{(1)}=0$ or~$y^{(2)}=0$ the assigned probabilities must differ, since
neither $y^{(1)}$ nor~$y^{(2)}$ can go negative.)

Just as in the ALOHAnet model, it~is natural to scale both time and and the
state space as the amount of memory~$N$ tends to infinity.  However,
we~shall not need to scale the model parameter~$p$.  Let ${\vec
x}=(x_1,x_2)=(y^{(1)},y^{(2)})/N$ be the normalized state of the
two-program system, and let $t=j/N$ be normalized time.  ${\vec x}$~jumps
at $t=1/N,2/N,3/N,\dotsc$ by~an amount $N^{-1}{\greekvec\xi}$, where
${\greekvec\xi}$~is a random variable with discrete density
\begin{equation}
\label{eq:constantdrift}
\Prob{\greekvec\xi={\vec z}} =
\cases{p/2,&if ${\vec z}=(1,0)$;\cr
       p/2,&if ${\vec z}=(0,1)$;\cr
       (1-p)/2,&if ${\vec z}=(-1,0)$;\cr
       (1-p)/2,&if ${\vec z}=(0,-1)$.\cr
	}
\end{equation}
As defined, the density of~${\greekvec\xi}$ is essentially independent
of~${\vec x}$.  It~is useful to relax this assumption, so~as to permit more
realistic stochastic modelling of dynamic data structures.
\begin{equation}
\label{eq:gendrift}
\Prob{\greekvec\xi({\vec x})={\vec z}} =
\cases{p(x_1)/2,&if ${\vec z}=(1,0)$;\cr
       p(x_2)/2,&if ${\vec z}=(0,1)$;\cr
       (1-p(x_1))/2,&if ${\vec z}=(-1,0)$;\cr
       (1-p(x_2))/2,&if ${\vec z}=(0,-1)$\cr
	}
\end{equation}
is a natural generalization.  Here $p(x)$ (assumed to take values between
$0$~and~$\frac12$ exclusive) specifies the probability of an increase in
memory usage by either program, as~a function of the fraction of available
memory which that program is currently using.  We~now write
${\greekvec\xi}$ as~${\greekvec\xi}({\vec x})$, to~indicate the dependence
of its density on~${\vec x}$.

The normalized state~${\vec x}$ is confined to the right triangle with
vertices $(0,0)$, $(1,0)$ and~$(0,1)$.  The expected drift
\begin{equation}
\langle{\greekvec\xi}({\vec x})\rangle =
\bigl(p(x_1)-{\textover12},p(x_2)-{\textover12}\bigr)
\end{equation}
may be viewed as a deterministic dynamical system on this two-dimensional
normalized state space.  Clearly, the vertex~$(0,0)$ is the global
attractor.  In~this model the two programs tend on the average not to use
much memory.

Since there is only a single attractor, the quantity of interest is the
expected time which must elapse before a fluctuation of specified magnitude
occurs.  Fluctuations which take the system state to the hypotenuse of the
triangle (where $x_1+x_2=1$, or~$y^{(1)}+y^{(2)}=N$) are {\em fatal\/}:
they correspond to memory exhaustion.  The rate at which they occur can be
estimated in the large-$N$ limit.

This is a two-dimensional system, so the optimal (least-action)
trajectories are not determined uniquely by the zero-energy constraint.
However we still have
\begin{equation}
\langle t_{\rm exit} \rangle \sim \exp (N{\cal S}_0), \qquad N\to\infty
\end{equation}
with ${\cal S}_0$ the action of the least-action trajectory which exits the
triangle through the hypotenuse.  The action is computed from the
Lagrangian dual to the Wentzell-Freidlin Hamiltonian
\begin{eqnarray}
\label{eq:stacksH}
H({\vec x},{\vec p}) & = &
\log\langle\exp({\vec p}\cdot{\greekvec\xi}({\vec x}))\rangle \\
& = &
 -\log2+\log\bigl\{\cosh p_x-[1-2p(x)]\sinh p_x \nonumber \\
&   & \qquad\qquad\quad  +\cosh p_y-[1-2p(y)]\sinh p_y\bigr\},
\nonumber
\end{eqnarray}
which follows from~(\ref{eq:gendrift}).

The zero-energy trajectories determined by~(\ref{eq:stacksH}) are studied
at length in Ref.~\cite{Maier6}, where it is shown that the MPEP depends
strongly on the behavior of the function~$p(x)$.  (See~Fig.~3.)  If~$p(x)$
is a strictly decreasing function, so~that the model is `increasingly
contractive,' with large excursions away from the attractor strongly
suppressed, then the MPEP turns out to be directed along the line segment from
$(0,0)$ to~$(\frac12,\frac12)$.  Its~action~is
\begin{equation}
{\cal S}_0 = 4\int_{x=0}^{1/2}\tanh^{-1}[1-2p(x)]\,dx.
\end{equation}
If on the other hand $p(x)$ is a strictly increasing function, so~that the
model is decreasingly contractive, with large excursions less strongly
suppressed, then there is a twofold degeneracy.  MPEPs are directed
outward from~$(0,0)$ to the two other vertices of the triangle, and
\begin{equation}
{\cal S}_0 = 2\int_{x=0}^{1}\tanh^{-1}[1-2p(x)]\,dx
\end{equation}
is their common action.

So when $p(x)$ is strictly increasing, there is a `hot spot' on the
hypotenuse of the triangle at~$(\frac12,\frac12)$.  When the two-program
system runs out of memory, as~$N\to\infty$ it is increasingly likely that
each program will be using approximately $N/2$~memory cells.  If~on the
other hand $p(x)$~is strictly decreasing, there are hotspots at the vertices
$(0,1)$
and~$(1,0)$.  Exhaustion increasingly tends to occur when one or the other
program is using all, or~nearly all, of~the $N$~memory cells.

If $p(x)$ is neither strictly increasing nor strictly decreasing, the
large-$N$ asymptotics may become more complicated.  The most easily treated
case is that of $p(x)\equiv p$, a~constant, \ie, the model
of~(\ref{eq:constantdrift}).  In~this model an {\em infinite degeneracy\/}
occurs: any~trajectory which moves some distance (possibly zero) from~$(0,0)$
toward $(0,1)$ or~$(1,0)$, and then moves into the interior of
the triangle at a $45^\circ$~angle until it~reaches the hypotenuse, is~a
least-action trajectory.  Large fluctuations away from the attractor may
proceed along any of this uncountable set of MPEPs.  As~a consequence there
is no hotspot: in~the large-$N$ limit, the exit location is uniformly
distributed over the hypotenuse.  Flajolet~\cite{Flajolet} first discovered
this phenomenon combinatorially, but it has a natural classical-mechanical
interpretation.  It~is however a bit counterintuitive: it~says that when
memory is exhausted, the fractions allocated to each program are as likely
to be small as large.  This is a very sensitive phenomenon.

\section{Conclusions}
We have seen that the Wentzell-Freidlin results on scaled jump processes
throw considerable light on the fluctuations of stochastically perturbed
dynamical systems, in~the weak-noise limit.  The appearance of a classical
Hamiltonian and Lagrangian, even if the unperturbed dynamical system is in
no sense Hamiltonian, is quite striking.  So~is the central importance of
zero-energy trajectories.

In this lecture I have focused on jump processes since they are the most
relevant to computer science applications.  (Computing is inherently
discrete.)  But they also occur in chemical physics: there is always an
integer number of molecules in any given region of space.  Attempts are now
being made to interpret the stochastic aspects of chemical reactions in
terms of optimal trajectories~\cite{Ross92}.  This is very reminiscent of
our focus on most probable exit paths~(MPEPs).

There is also a large deviation theory of continuous-time
processes~\cite{VF,Varadhan}, such as the diffusion processes specified by
the stochastic differential equation~(\ref{eq:stochDE}).  Associated to
each such process is a Fokker-Planck equation (a~parabolic partial
differential equation) describing the diffusion of probability.  The
zero-energy classical trajectories of continuous-time large deviation
theory can be viewed as the {\em characteristics\/} of this differential
equation.  Normally one expects only hyperbolic equations to have
characteristics, but these characteristics are emergent: they manifest
themselves only in the weak-noise limit.

A large deviation theory of spatially extended systems would be an
interesting extension, but is still under development.  Such systems
include stochastic partial differential equations and stochastic cellular
automata.  In~such systems a MPEP would be a trajectory in the system state
space, describing a most probable {\em spatially extended\/} fluctuation
leading from one metastable state to another.  Much work has been done on
this by statistical mechanicians and field theorists (who~call such
fluctuations `instantons'~\cite{Schulman}), but the theory is less
complete than the theory I~have sketched in this lecture.  The theory of
extended fluctuations has in~particular not been applied to distributed
computer systems.  There is clearly much left to be done!
\vfill\eject

\vfill\eject

\begin{figure}
\caption{The expected drift velocity $\langle\xi(x)\rangle$ of the
stochastic ALOHAnet model, as a~function of normalized network state~$x$.
Model parameters are $q_0=0.43$ and~$q_1=5.0$, as
in Ref.~\protect\cite{Gunther2}.}
\label{fig:1}
\end{figure}

\begin{figure}
\caption{An impressionistic sketch of the parameter space of the
stochastic ALOHAnet model.  Within the horn-shaped region the network
is bistable; outside~it, monostable.  The tip of the horn is analogous to
a statistical-mechanical critical point.}
\label{fig:2}
\end{figure}

\begin{figure}[t]
\caption{The triangular normalized state space of the colliding stacks
model.  Trajectory~T1 is the most probable exit path when the
function~$p(x)$ is strictly decreasing, but if $p(x)$ is strictly
increasing then T2~and~T2$^\prime$ are both MPEPs.
Trajectory~T3 is one of the uncountably many MPEPs which arise
when $p(x)$ is independent of~$x$.}
\label{fig:3}
\end{figure}
\vfill\eject


{\small
\begin{thebibliography}{10}
\bibitem{Bertsekas87}
D.~Bertsekas and R.~Gallager.
\newblock {\em Data Networks}.
\newblock Prentice-Hall, Englewood Cliffs, New Jersey, 1987.

\bibitem{Chinarov93}
V.~A. Chinarov, M.~I. Dykman, and V.~N. Smelyanskiy.
\newblock Dissipative Corrections to Escape Probabilities of Thermally
  Nonequilibrium Systems.
\newblock Preprint.

\bibitem{Flajolet}
P.~Flajolet.
\newblock The Evolution of Two Stacks in Bounded Space and Random Walks in a
  Triangle.
\newblock In {\em Mathematical Foundations of Computer Science: Proceedings of
  the 12th Symposium\/} (Bratislava, Czechoslovakia, 1986), pp.\ 325--340.
 Springer-Verlag, New York/Berlin, 1986.

\bibitem{VF}
M.~I. Freidlin and A.~D. Wentzell.
\newblock {\em Random Perturbations of Dynamical Systems}.
\newblock Springer-Verlag, New York/Berlin, 1984.

\bibitem{Gunther2}
N.~J. G{\"u}nther and J.~G. Shaw.
\newblock Path Integral Evaluation of {ALOHA} Network Transients.
\newblock {\em Inf.\ Process.\ Lett.} {\bf 33} (1990), 289--295.

\bibitem{Hanggi90}
P.~H{\"a}nggi, P.~Talkner, and M.~Borkovec.
\newblock Reaction-rate Theory: Fifty Years After {Kramers}.
\newblock {\em Rev.\ Modern Phys.} {\bf 62} (1990), 251--341.

\bibitem{Huberman88}
B.~A. Huberman, editor.
\newblock {\em The Ecology of Computation}.
\newblock Elsevier, New York/Amsterdam, 1988.

\bibitem{Lande85}
R.~Lande.
\newblock Expected Time for Random Genetic Drift of a Population Between Stable
  Phenotypic States.
\newblock {\em Proc.\ Nat.\ Acad.\ Sci.\ USA\/} {\bf 82} (1985), 7641--7645.

\bibitem{Lim87}
J.-T. Lim and S.~M. Meerkov.
\newblock Theory of Markovian Access to Collision Channels.
\newblock {\em {IEEE} Trans.\ Comm.} {\bf COM-35} (1987), 1278--1288.

\bibitem{Louchardprime}
G.~Louchard and R.~Schott.
\newblock Probabilistic Analysis of Some Distributed Algorithms.
\newblock {\em Random Structures and Algorithms\/} {\bf 2} (1991), 151--186.
\newblock A preliminary version appeared in {{\em Proceedings of CAAP~'90\/}},
  Lecture Notes in Computer Science {\#431}, pp.\ 177--190. Springer-Verlag.

\bibitem{Maier6}
R.~S. Maier.
\newblock Colliding Stacks: A Large Deviations Analysis.
\newblock {\em Random Structures and Algorithms\/} {\bf 2} (1991), 379--420.

\bibitem{Maier3}
R.~S. Maier.
\newblock A Path Integral Approach to Data Structure Evolution.
\newblock {\em J.~Complexity\/} {\bf 7} (1991), 232--260.

\bibitem{Maier7b}
R.~S. Maier.
\newblock Communications Networks as Stochastically Perturbed Nonlinear
  Systems: A~Cautionary Note.
\newblock In {\em Proceedings of the 30th {Allerton} Conference on
  Communication, Control and Computing}, pp.\ 674--681, Monticello, Illinois,
  1992.

\bibitem{MaierA}
R.~S. Maier and D.~L. Stein.
\newblock Transition-rate Theory for Non-Gradient Drift Fields.
\newblock {\em Phys.\ Rev.\ Lett.} {\bf 69} (1992), 3691--3695.

\bibitem{MaierB}
R.~S. Maier and D.~L. Stein.
\newblock The Escape Problem for Irreversible Systems.
\newblock Submitted to {{\em Physical Review~E\/}}, 1993.

\bibitem{Matkowsky84}
B.~J. Matkowsky, Z.~Schuss, C.~Knessl, C.~Tier, and M.~Mangel.
\newblock Asymptotic Solution of the {Kramers-Moyal} Equation and First-Passage
  Times for {Markov} Jump Processes.
\newblock {\em Phys.\ Rev.~A\/} {\bf 29} (1984), 3359--3369.

\bibitem{Naeh90}
T.~Naeh, M.~M. K{\l}osek, B.~J. Matkowsky, and Z.~Schuss.
\newblock A Direct Approach to the Exit Problem.
\newblock {\em {SIAM} J.~Appl.\ Math.} {\bf 50} (1990), 595--627.

\bibitem{Nelson84}
R.~Nelson.
\newblock The Stochastic Cusp, Swallowtail, and Hyperbolic Umbilic Catastrophes
  as Manifest in a Simple Communications Model.
\newblock In {\em {Performance} '84}, edited by E.~Gelenbe, pp.\ 207--224.
  North-Holland, 1984.

\bibitem{Nelson}
R.~Nelson.
\newblock Stochastic Catastrophe Theory in Computer Performance Modeling.
\newblock {\em J.~Assoc.\ Comput.\ Mach.} {\bf 34} (1987), 661--685.

\bibitem{Newman85}
C.~M. Newman, J.~E. Cohen, and C.~Kipnis.
\newblock Neo-{Darwinian} Evolution Implies Punctuated Equilibria.
\newblock {\em Nature} {\bf 315} (1985), 400--401.

\bibitem{Ross92}
J.~Ross, K.~L.~C. Hunt, and P.~M. Hunt.
\newblock Thermodynamic and Stochastic Theory for Nonequilibrium Systems with
  Multiple Reactive Intermediaries.
\newblock {\em J.~Chem.\ Phys.} {\bf 96} (1992), 618--629.

\bibitem{Schulman}
L.~S. Schulman.
\newblock {\em Techniques and Applications of Path Integration}.
\newblock Wiley, New York, 1981.

\bibitem{Schuss80}
Z.~Schuss.
\newblock {\em Theory and Application of Stochastic Differential Equations}.
\newblock Wiley, New York, 1980.

\bibitem{Stallings88}
W.~Stallings.
\newblock {\em Data and Computer Communications}.
\newblock Macmillan, New York, second edition, 1988.

\bibitem{vanKampen81}
N.~G. van Kampen.
\newblock {\em Stochastic Processes in Physics and Chemistry}.
\newblock North-Holland, New York/Amsterdam, 1981.

\bibitem{Varadhan}
S.~R.~S. Varadhan.
\newblock {\em Large Deviations and Applications}.
\newblock Society for Industrial and Applied Mathematics, Philadelphia, 1984.

\bibitem{Wentzell76}
A.~D. Wentzell.
\newblock Rough Limit Theorems for Large Deviations for {Markov} Processes.
\newblock {\em Theory Probab.\ Appl.} {\bf 21} (1976), 227--242, 499--512.
\end{thebibliography}}
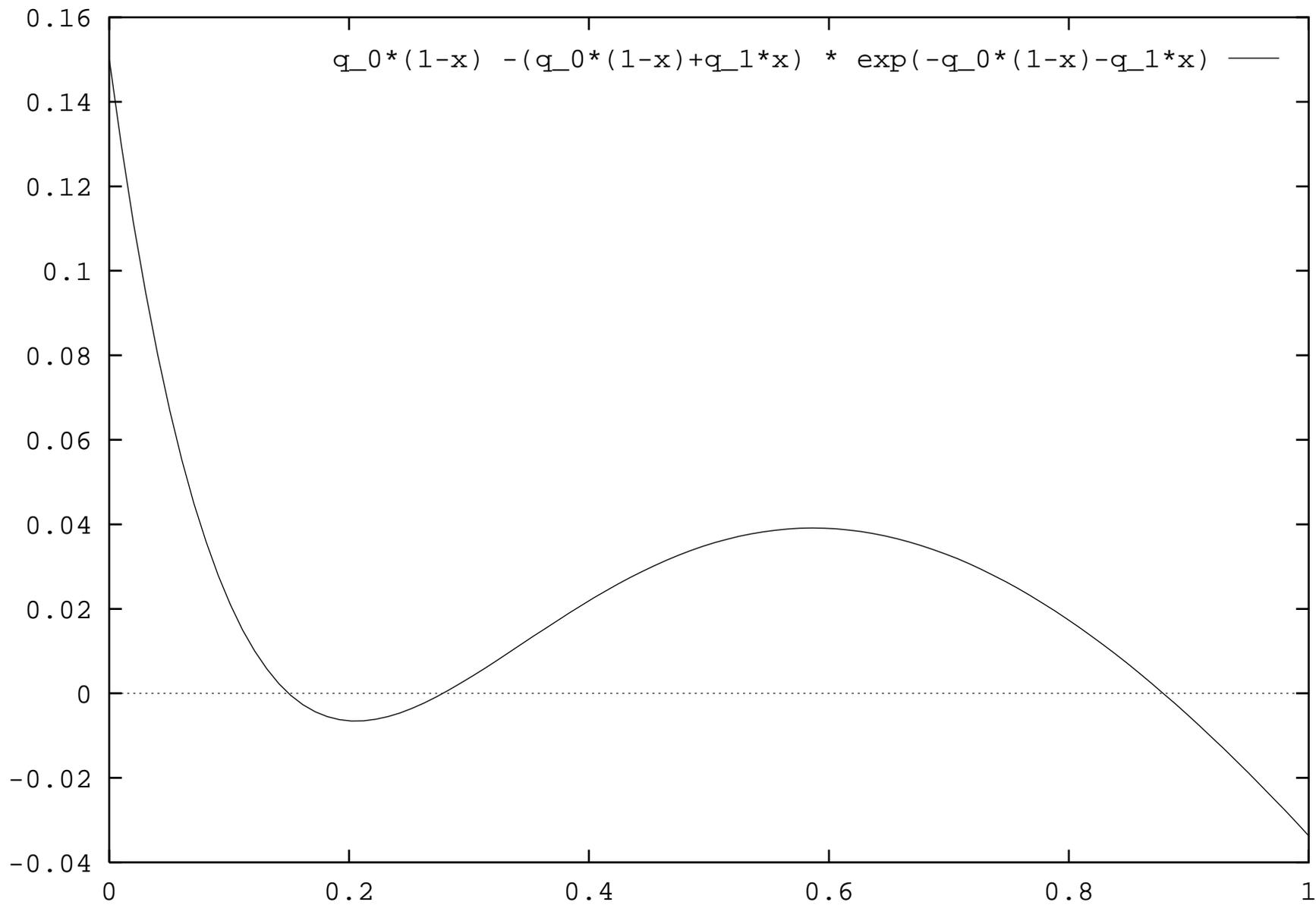
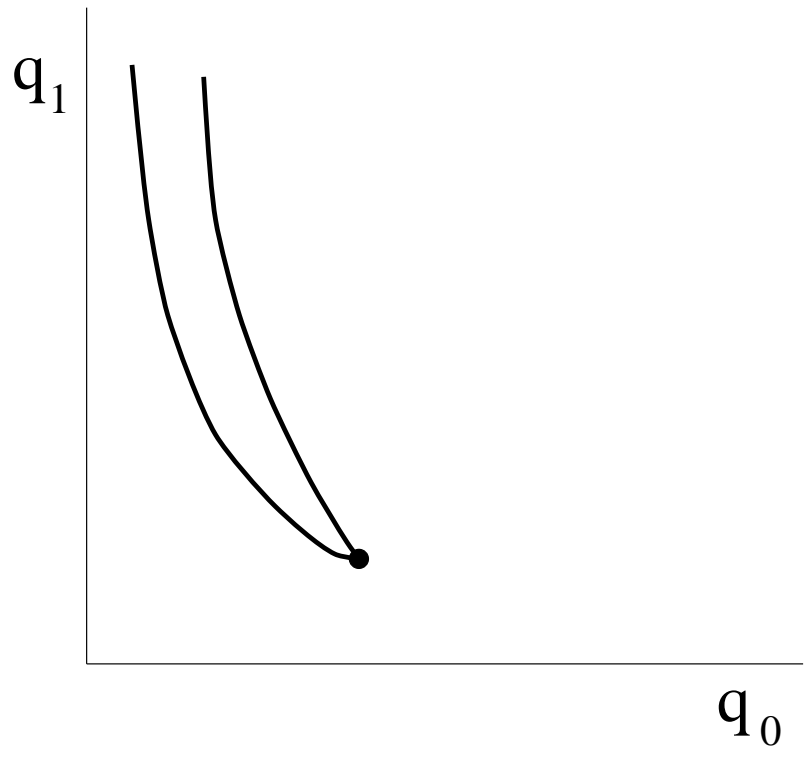
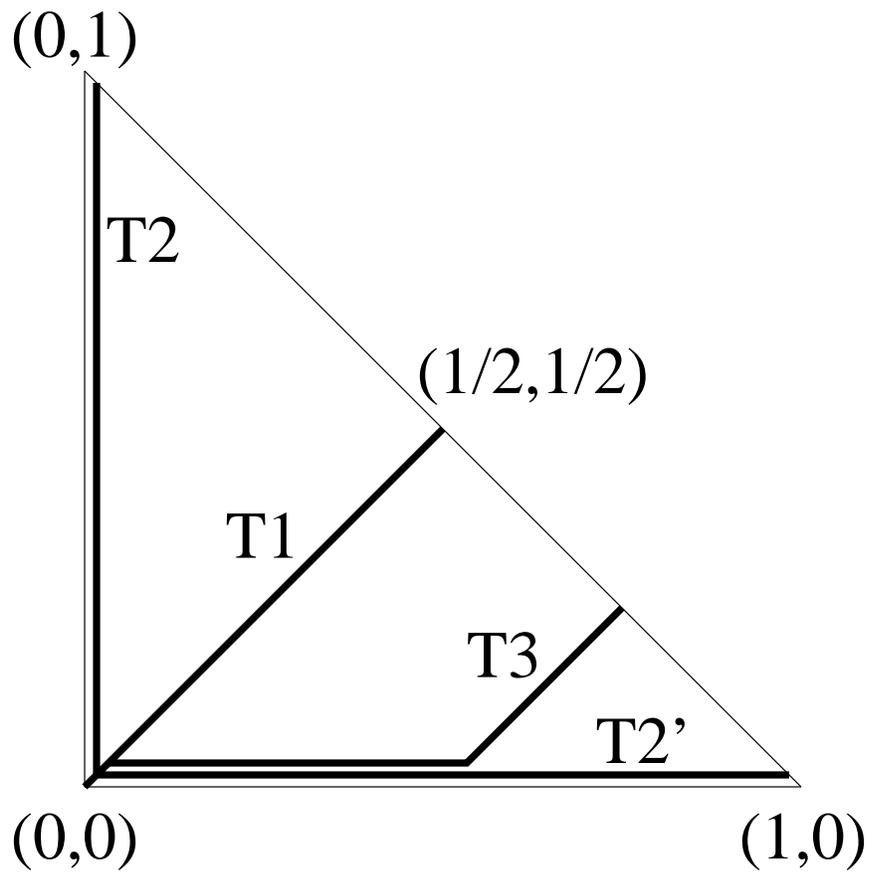
\end{document}